\documentclass[journal]{IEEEtran}
\usepackage{color}
\usepackage{cite}
\usepackage{graphics} 
\usepackage{epsfig} 
\usepackage{times} 
\usepackage{amsmath} 
\usepackage{amssymb}  
\usepackage{subfigure}
\usepackage{url}
\newtheorem{theorem}{Theorem}

\usepackage{color}
\usepackage{algorithm}

\newtheorem{definition}[theorem]{Definition}

{ 

}

\newcommand{\bi}{\begin{itemize}}
\newcommand{\ei}{\end{itemize}}
\newcommand{\bd}{\begin{displaymath}}
\newcommand{\ed}{\end{displaymath}}
\newcommand{\be}{\begin{eqnarray*}}
\newcommand{\ee}{\end{eqnarray*}}

\ifCLASSINFOpdf

\else

\fi

\begin{document}
\title{\Large Data-driven Identification and Prediction of Power System Dynamics Using Linear Operators}
\author{\IEEEauthorblockN{Pranav Sharma\IEEEauthorrefmark{1}, Bowen Huang\IEEEauthorrefmark{2}, Umesh Vaidya\IEEEauthorrefmark{3}, Venkatramana Ajjarapu\IEEEauthorrefmark{4}}\\
\IEEEauthorblockA{Department of Electrical and Computer Engineering\\
Iowa State University, Ames, Iowa, USA\\
Email: \IEEEauthorrefmark{1}spranav@iastate.edu, \IEEEauthorrefmark{2}bowen@iastate.edu,
\IEEEauthorrefmark{3}ugvaidya@iastate.edu,
\IEEEauthorrefmark{4}vajjarap@iastate.edu }
\thanks{ Financial support from the Department of Energy’s Grant number DE-OE 0000876 is gratefully acknowledged.}}

\maketitle

\begin{abstract}

 In this paper, we propose linear operator theoretic framework involving Koopman operator for the data-driven identification of power system dynamics. We explicitly account for noise in the time series measurement data and propose robust approach for data-driven approximation of Koopman operator for the identification of nonlinear power system dynamics. The identified model is used for the prediction of state trajectories in the power system. The application of the framework is  illustrated using an IEEE nine bus test system.
\end{abstract}

\IEEEpeerreviewmaketitle
\section{Introduction}\label{section_intro}

The advent of synchrophasor measurements in power system has contributed towards high temporal and spatial granularity in Wide Area Measurement Systems(WAMS). PMUs provide wide-area visualization, state estimation and voltage monitoring for monitoring power system operation. Due to the high granularity of PMU data, challenges  related to system dynamics such as oscillation detection, stability analysis and control can also be addressed \cite{uhlen2016synchrophasor, schweitzer2008advanced, de2010synchronized}.
Various data-driven methods for dynamic behavior identification have been developed using synchrophasor measurements, methods based on subspace identification and prony analysis have also been developed \cite{yin2015data,larsson2009monitoring, messina2006interpretation, liu2008oscillation}.
\footnote{$\textcopyright$ 2019 IEEE. Personal use of this material is permitted. Permission from IEEE must be obtained for all other uses, in any current or future media, including reprinting/republishing this material for advertising or promotional purposes, creating new collective works, for resale or redistribution to servers or lists, or reuse of any copyrighted component of this work in other works.}

While being able to address key aspects of dynamic identification, these methods have dependency on knowledge of system model and more importantly, assume a linear dynamic system. In this context, linear operator based method stands out as it provide a way to capture the nonlinear dynamics in a model free scenario \cite{susuki2011nonlinear,susuki2014nonlinear,susuki2016applied}. In particular, Koopman operator based dynamic mode decomposition (DMD) method has been developed for modal identification \cite{budivsic2012applied, williams2015data}. These methods find applications in identifying inter-area oscillation\cite{barocio2015dynamic}, partitioning power network\cite{raak2016data}, identifying model parameters \cite{susuki2018estimation} and stability analysis \cite{mauroy2016global}. These linear operator based techniques provide a great insight in system behavior in a model free scenario. However, these methodologies are highly sensitive to the data quality.

{As given in IEEE standards on synchrophasors \cite{martin2008exploring}, the upper bound on signal to noise ratio (SNR) for PMU measurements is 20dB for steady state (Smaller the SNR, higher is the error) . Although, there are no such standards for noise in measurements when system is undergoing dynamics. However, practical studies suggest that dynamic measurements often have SNR in the range of 30-20dB \cite{phadke2009synchronized, becejac2016analysis}. Thus it is important for any data driven technique to take into account the ambient noise present in measurements. Furthermore, an imperfect communication infrastructure for WAMS can lead to missing data and bad data in measurements.}

In the previous work of the second author\cite{sinha2018robust}, we have shown that existing DMD methods fail to capture true system dynamics in presence of measurement and process noises . As previously stated, linear operator method provides a model-free framework for system identification which means there is no external way to justify the accuracy of data under consideration. Thus it becomes important to have a framework for system identification that is robust to various process and measurement noises present in the real world scenario and can identify underlying dynamic system accurately.

Our objective is to provide a robust framework for model free data-driven identification of system dynamics. The novelty of our method lies in the robustness of our algorithm, which takes into account noise, missing data and outliers. In a realistic scenario it is important to account for various noises associated with wide area measurements, that can affect the identification and prediction capability of any algorithm. 

For a data driven model-free technique it is important to identify underlying dynamic system accurately. One validation of any data driven method is how accurately can it predict the future evolution of given measurements. Similar to system identification, various techniques for dynamic state estimation have been developed using Kalman filter and extended Kalman filter states\cite{singh2014decentralized}. Similar to existing method of dynamic system identification these methods have dependency of model information and holds good for linear dynamics. Recent work has used Koopman operator based Kalman filter for trajectory prediction, that address the issue of model dependency and nonlinear  \cite{netto2018robust}. However, as pointed out earlier these methods are susceptible to data quality. Thus one of the key objective of this work is to provide a robust framework for model free data driven trajectory prediction for nonlinear dynamic system. 

To this end we have developed a robust approximation of Koopman operator that addresses the challenges associated with existing data driven system identification techniques, as highlighted above \cite{2018arXiv180308562S}. 

 Thus the overall objective of our work is to provide a robust framework for dynamic system identification and prediction in a  realistic power system scenario. The main contributions of the paper are the  following: We propose robust approach for the approximation of the Koopman operator that explicitly accounts for the noise in the measurement data. The robust approximation procedure is used for the robust identification of nonlinear power system dynamics using noisy data set. The robust predictor is used for the estimation of power system dynamics. Impact of training data length used in the approximation of Koopman operator on the prediction error is also analyzed.

 The rest of the paper is organized as follows. In section \ref{section_preliminaries}, we discuss our methodology for robust computation of Koopman operator and trajectory prediction. Section \ref{section_power_case}, illustrates the details of power system test case under consideration and system properties. Section \ref{section_sensitivity_of_robust_framework}, presents detailed simulation studies for the given test system for various realistic scenarios of power system measurement. The conclusion and contribution of our work is illustrated in section \ref{section_conclusion}, along with future implications of this work.

\section{Proposed Method}\label{section_preliminaries}
\subsection{Transfer operator for stochastic system}
For a random dynamic system of form:
\begin{eqnarray}\label{eqn_rds}
x_{t+1}=T(x_t,\xi_t)\label{system}
\end{eqnarray}
where $T:X\subset W \to  X$ with $X\subset \mathbb{R}^N$ is assumed to be invertible with respect to $x$ for each fixed value of $\xi$. $\xi_t\in W$ is an independent variable $\vartheta$ i.e., 
\[{\rm Prob}(\xi_t\in B)=\vartheta(B)\]
for every set $B\subset W$ and all $t$. For such a discrete dynamical system we can define Koopman linear operator, as follows:
\begin{definition} [Koopman Operator] Given any $h\in\cal{F}$, $\mathbb{U}:{\cal F}\to {\cal F}$ is defined by
\[[\mathbb{U} h](x)={\bf E}_{\xi}[h(T(x,\xi))]=\int_W h(T(x,v))d\vartheta(v)\]
\end{definition}

\subsection{Robust approximation of Koopam operator}
In this section, the robust approximation algorithm of Koopman operator  will be combined with the power system settings, which is proposed in the earlier work\cite{2018arXiv180308562S}.

Consider snapshots of state variables data set for power system in form of equation (\ref{eqn_rds}),
\begin{eqnarray}
 X = [x_0,x_2,\ldots,x_M]
 \label{data}
\end{eqnarray}
where $x_i\in X\subset \mathbb{R}^n$. The data-set $\{x_k\}$ is state trajectory subjected to various processes and measurement noises. 

In particular, we consider norm bounded uncertainty in the data set. Since the trajectory $\{x_k\}$ is one particular realization of the RDS, the other random realization can be assumed to be obtained by perturbing $\{x_k\}$. We assume that the data points $x_k$ are perturbed by norm bounded deterministic perturbation of the form 
\[\delta x_k=x_k+\delta,\;\;\; \delta\in \Delta.\]

We define $\mathcal{D}=\{\psi_1,\psi_2,\ldots,\psi_K\}$ as the observables on $x_k$. These observables belong to $\psi_i\in L_2(X,{\cal B},\mu)={\cal G}$, where $\mu$ is some positive measure. Let ${\cal G}_{\cal D}$ denote the span of ${\cal D}$ such that ${\cal G}_{\cal D}\subset {\cal G}$. The chosen observables should be rich enough to approximate the leading eigenfunctions of Koopman operator. Define vector valued function $\mathbf{\Psi}:X\to \mathbb{C}^{K}$ as
\begin{equation}
\mathbf{\Psi}(x):=\begin{bmatrix}\psi_1(x) & \psi_2(x) & \cdots & \psi_K(x)\end{bmatrix}\label{dic_function}
\end{equation}

Here, $\mathbf{\Psi}$ lift the system from state space to feature space. Any function $\phi,\hat{\phi}\in \mathcal{G}_{\cal D}$ can be written as
\begin{eqnarray}
\phi = \sum_{k=1}^K a_k\psi_k=\boldsymbol{\Psi a},\quad \hat{\phi} = \sum_{k=1}^K \hat{a}_k\psi_k=\boldsymbol{\Psi \hat{a}}\label{expand}
\end{eqnarray}
for some set of coefficients $\boldsymbol{a},\boldsymbol{\hat{a}}\in \mathbb{C}^K$. 
Let \begin{eqnarray}
 \hat{\phi}(x)=[\mathbb{U}\phi](x)+r=E_\xi[\phi(T(x,\xi))]+r. \label{residual}
\end{eqnarray}

Different realization of the system will be of form $\{x_k+\delta\}$ with $\delta\in \Delta$ to write (\ref{residual}) as follows:
\begin{eqnarray}
\hat{\phi}(x_m + \delta x_m)=\phi(x_{m+1})+r, \;\;\;k=1,\ldots,M-1.
\end{eqnarray}
Here the objective is to minimize this residue for all possible pair of data points of the form $\{x_m+\delta,x_{m+1}\}$. 
Using (\ref{expand}) we write the above as follows:
\[
\boldsymbol{\Psi}(x_k + \delta x_k)\boldsymbol {\hat{a}}=\boldsymbol{\Psi}(x_{k+1})\boldsymbol {a}+r.
\]

Our aim is to find $\bf K$, finite approximation of Koopman operator that maps $\boldsymbol{a}$ to $\boldsymbol{\hat{a}}$, i.e., ${\bf K}\boldsymbol{a}=\boldsymbol{\hat a}$, and minimize the the residue term, $r$.  Multiplying by $\boldsymbol{\Psi}^\top( x_m)$ on both the sides of above expression and summing over $m$ we obtain 
\[\left[\frac{1}{M}\sum_m \boldsymbol{\Psi}^\top( x_m) \boldsymbol{\Psi}(x_m + \delta x_m){\bf K}-\boldsymbol{\Psi}^\top(x_m)\boldsymbol{\Psi}( x_{m+1})\right]{\boldsymbol a}.\]
For robust approximation, uncertainty penalizes the estimation. Thus, robust optimization can be written as a $\min-\max$ convex optimization problem, as follow: 

\begin{equation}\label{edmd_robust_convex}
\min\limits_{\bf K}\max_{\delta{\bf G}\in \bar \Delta}\parallel ({\bf G}+\delta {\bf G}){\bf K}-{\bf A}\parallel_F
\end{equation}
where ${\bf K},{\bf G}_\delta,{\bf A}\in\mathbb{C}^{K\times K}$ and $\delta {\bf G}\in \mathbb{R}^{K\times K}$ is the new perturbation term characterized by uncertainty set  $\bar \Delta$ which lies in the feature space of dictionary function and the matrix ${\bf G}=\frac{1}{M}\sum_{m=1}^M \boldsymbol{\Psi}({ x}_m)^\top \boldsymbol{\Psi}({x}_m)$. $\bar \Delta$ is the new uncertainty set defined in the feature space and will inherit the structure from set $\Delta $ in the data space.

The equivalence between the robust optimization problem (\ref{edmd_robust_convex}) and $\ell_1$ Lasso regularization can be established as follows.
\begin{theorem}
Defining
\[\bar \Delta:=\{{\delta \bf G }=(\delta {\bf G}_1,\ldots, \delta{\bf G}_K)\in \mathbb{R}^{K\times K}: \parallel \delta{\bf G}_i\parallel_2\leq c\}.\]
Following two optimization problems are equivalent
\begin{eqnarray}
&& \min\limits_{\bf K}\max_{\delta {\bf G}\in \bar \Delta}\parallel ({\bf G}+\delta {\bf G}){\bf K}-{\bf A}\parallel_F \\
&& \min\limits_{\bf K}\parallel {\bf G}{\bf K}-{\bf A}\parallel_F+c \sum_{k=1}^K\parallel {\bf K}_k\parallel_1\label{regular1}
\end{eqnarray}
where ${\bf K}_k $ is the $k^{th}$ column of matrix $\bf K$. 
\end{theorem}
The complete boundness proof and equivalence proof can be found in \cite{sinha2018robust,caramanis201214}.

\subsection{Design of robust predictor in power system}\label{subsection_predictor_design}

Based on the robust optimization formulation (\ref{edmd_robust_convex}) and the Lasso type regularization term in (\ref{regular1}), it provides a systematic way of tuning the regularization parameter. The conception of the optimization problem finds a parallelism in the machine learning area as well, involving the over-fitting and under-fitting problem.
Here regularization term implies that the operator framework can fit into data driven predictor design.
We first approximate the transfer operator using training data. Let $\{x_0,\ldots,x_M\}$ be the training data-set and $\bf K$  be the finite-dimensional approximation of the robust Koopman operator (\ref{edmd_robust_convex}). 
Let $\bar x_0$ be the starting point for trajectory prediction.

The initial condition from state space is mapped to the feature space, i.e., 
\[\bar x_0\implies {\bf \Psi}(\bar x_0)^\top=: {\bf z}\in \mathbb{R}^K.\] Using Koopman operator system propagates as \[{\bf z}_n={\bf K}^n{\bf z}.\]
From this we can obtain the trajectory in state space, as
\[\bar x_n=C {\bf z}_n\]
where matrix $C$ is obtained as the solution of the following least square problem
\begin{eqnarray}\label{C_pred}
\min_C\sum_{i = 1}^M \parallel x_i - C \boldsymbol \Psi (x_i)\parallel_2^2
\end{eqnarray}
This notion of robust approximation of Koopman operator and trajectory prediction will be used for power system dynamic identification and prediction. 

\section{Power System Test Case}\label{section_power_case}
In order to understand the implications of developed robust prediction and identification algorithm for power system, we consider IEEE 9 bus test case. For the given 9 bus system, synthetic data is generated using detailed modeling of generator, exciter and governor dynamics at the generator buses. 

\begin{figure}[htp!]
\centering
\includegraphics[scale=.3]{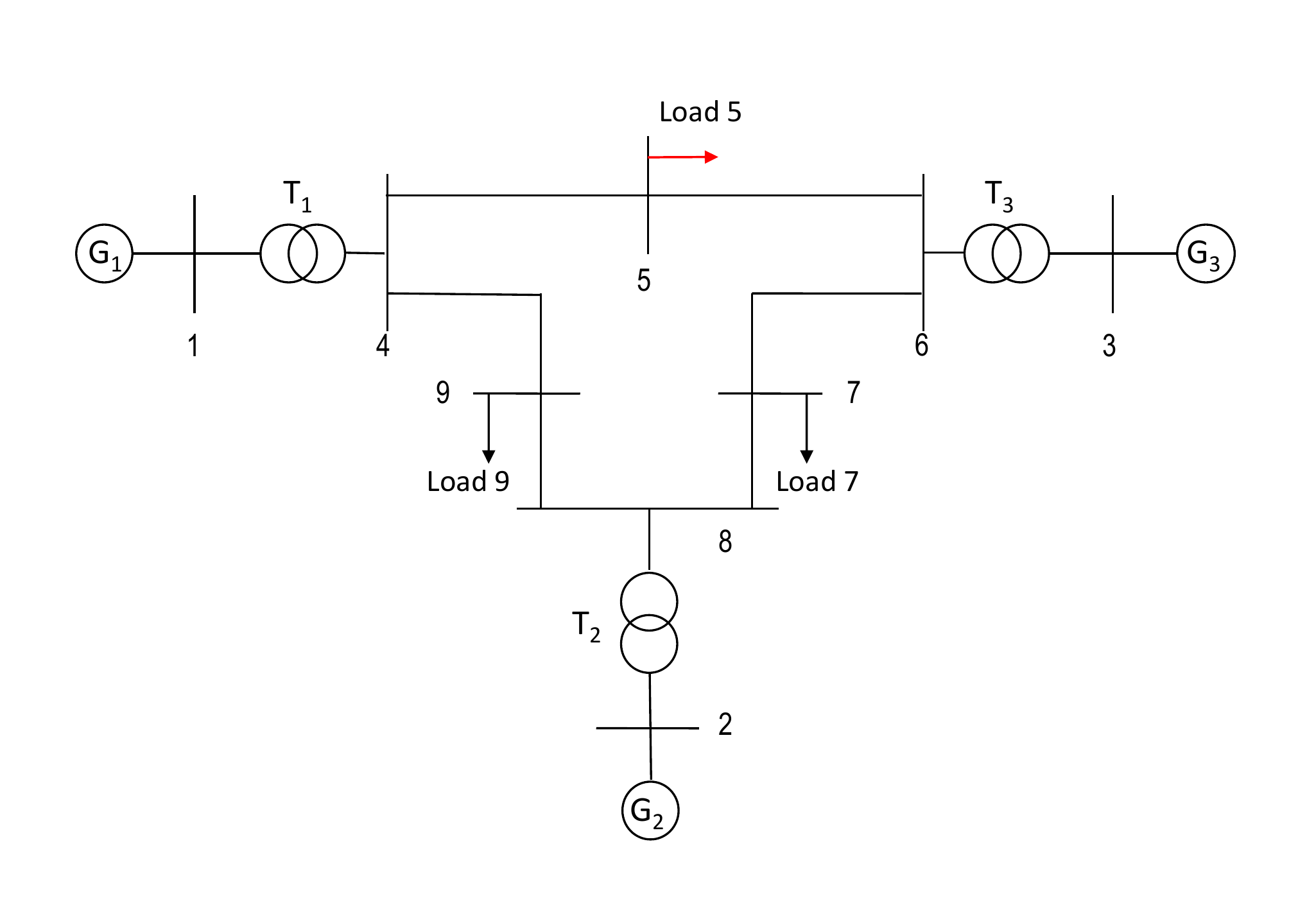}
\caption{IEEE 9 bus system}\label{figure_9bus_diag}
\end{figure}

 For the given system a $5^{th}$ order generator dynamic model is considered, details of the model can be found in \cite{Sauer_pai_book}. Each generator has an IEEE type-I exciter with $4^{th}$ order dynamic model along with a $1^{st}$ order governor control. The detailed discussion on the DAE model under consideration is presented in {\cite{khaitan2013high}} and omitted from this paper. In brief, the system can be described as follow:
\begin{itemize}
    \item \textit{Generator Model:} A $5^{th}$ order generator dynamic model is considered with genrator states $\delta, \omega_g, E_q^{'}, E_{d_1}^{'} E_{d_2}^{'} $
    \item \textit{Exciter Model:} IEEE Type I exciter is considered for each generator consisting of a $4^{th}$ order dynamic model.
    \item \textit{Governor Model:} A type II governor model is considered for each generator with first order dynamics. 
 \end{itemize}

\begin{figure}[htp!]
\centering
\includegraphics[scale=.7]{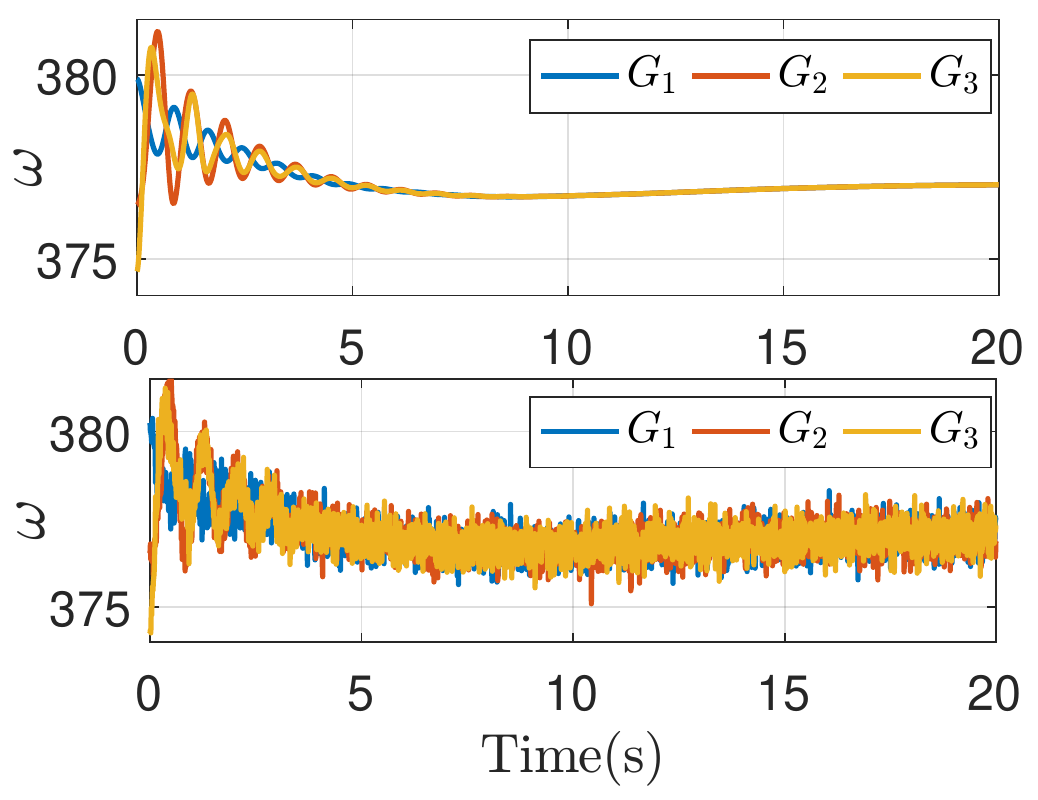}
\caption{Post fault frequency measurement (a) Purely deterministic measurement (b) Measurement corrupted with ambient noise of 20 dB}\label{fig_9bus_timeplot}
\end{figure}

Thus the dynamic model as represented in (\ref{system}) has $x \in \mathbb{C}^{30}$, here these state variables are measured as a response to a three phase fault at a load bus. For a given fault, system oscillates and settles at the equilibrium point, as shown in figure \ref{fig_9bus_timeplot}(a) in terms of generator angular speed ($\omega $). Here angular speed is calculated as follows. 
\begin{equation*}
    \omega = 2 \pi \cdot (60) *  \omega_g (p.u.)
\end{equation*}

For the given system, synthetic data is generated using PSAT toolbox \cite{milano2005open} for MATLAB. State measurements are recorded for multiple fault scenarios, load variation and system perturbation with sampling frequency of $0.01 sec$ which is in accordance with PMU measurements. Further, measurement noise in form of white Gaussian noise is introduced in the generated data, as shown in figure \ref{fig_9bus_timeplot}(b). At a given operating point we can obtain a linearized dynamic model around which system oscillates, the modes of this linearized system are compared with the identified dynamic system. The first test of our method is to identify the dominant modes of the underlying dynamic system using these measurements. Here we recorded measurements for all states of the dynamic system, as they evolve after the fault. In addition to this our robust predictor is used for trajectory prediction for the given measurements. 

\section{Simulation Studies}\label{section_sensitivity_of_robust_framework}
\subsection{Identification and prediction under ideal condition}
\begin{figure}[htp!]
\centering
\includegraphics[scale=.6]{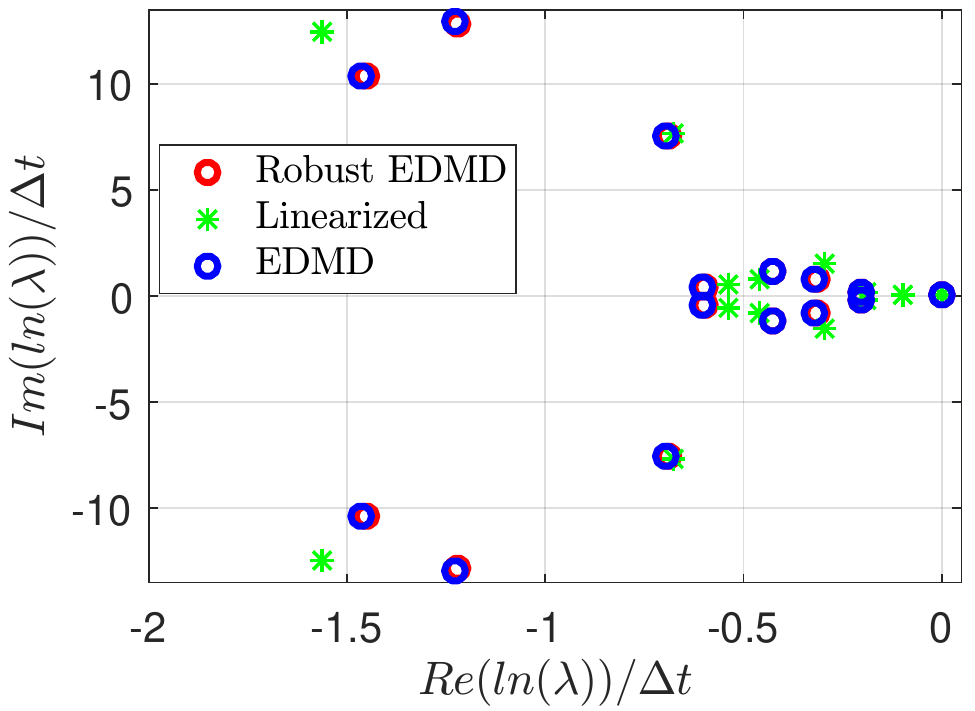}
\caption{System mode identification using Robust EDMD and EDMD for deterministic measurement}\label{fig_no_noise}
\end{figure}

Robust EDMD method performs as good as existing EDMD method for a deterministic system without any noise. Measurements recorded for a deterministic system as shown in figure \ref{fig_9bus_timeplot}(a), system modes are identified using EDMD and Robust EDMD algorithm, as shown in figure \ref{fig_no_noise}. 

It is important to highlight that, data driven system mode identification can be as good as the data itself. That is, if the dynamic system is not excited enough then it is impossible to identify all the modes from the given data. This persistence of excitation is critical in all data driven identification methods; interested readers are referred to \cite{willems2005note} for a detailed discussion. Therefore, to check the accuracy of further analysis with noise and variable data length, system identification in ideal condition is considered as a reference.  

\subsection{Effect of measurement noise}

\begin{figure}[htp!]
\centering
\subfigure[]{\includegraphics[scale=.52]{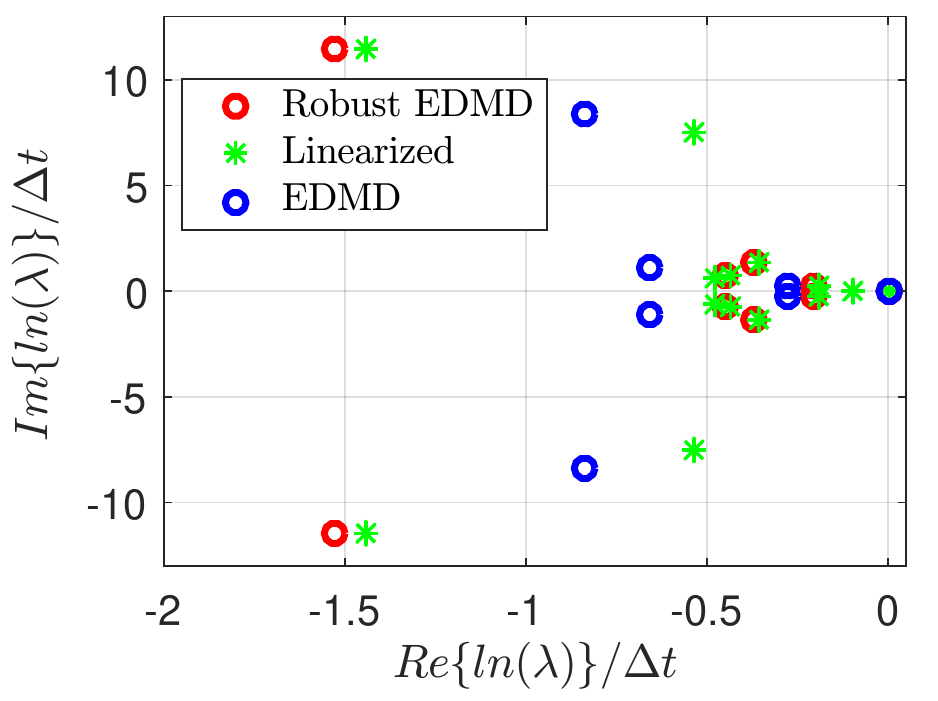}}
\subfigure[]{\includegraphics[scale=.52]{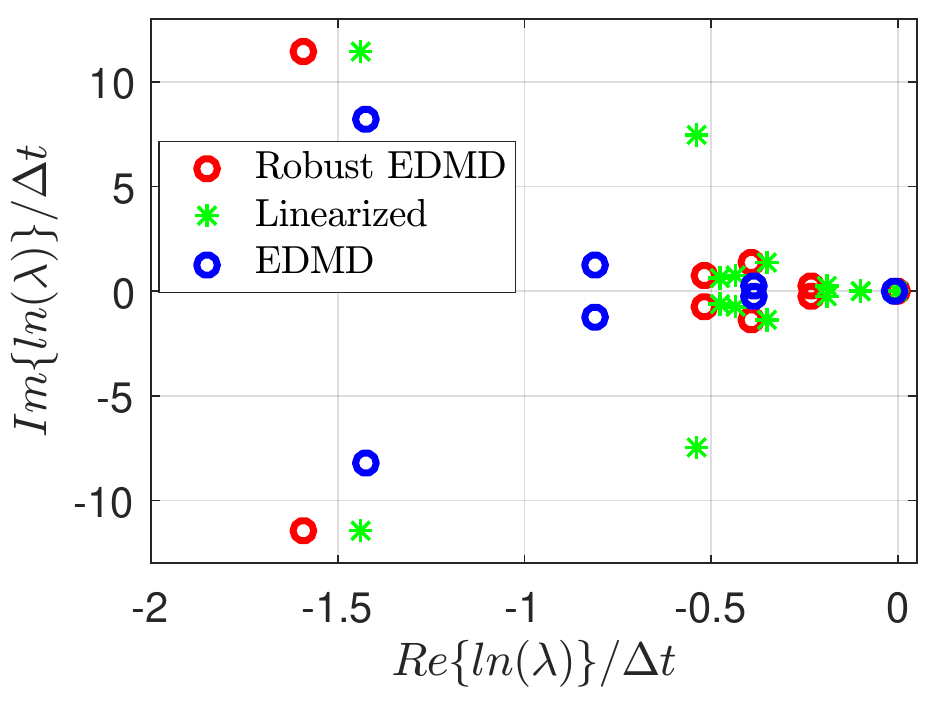}}
\caption{ System mode identification using Robust EDMD and EDMD (a) For ambient noise of SNR 20 dB (b) For ambient noise of SNR 17 dB}\label{fig_noise_level}
\end{figure}

State measurements corrupted with noise are considered for system identification, as shown in figure \ref{fig_9bus_timeplot}(b). For such measurements, robust linear operator identifies dominant modes with great accuracy. As shown in figure \ref{fig_noise_level}, two scenarios with noise level 20dB and 17 dB are considered. Robust EDMD identifies dominant modes closer to $j\omega$ axis, where existing method of dynamic mode decomposition fails to identify the system. 

\subsection{Trajectory Prediction} As illustrated in section \ref{subsection_predictor_design}, robust predictor can be designed to predict the evolution of dynamic measurements. To this end we take a rolling window of 4 seconds to predict system evolution for next 2 seconds. Though it is intuitive that the prediction accuracy will increase with increase in data sample. However in this case, as power system quickly settles down after clearing the fault, it is important to predict the transient behavior of system states, immediately after the fault. 
\begin{figure}[htp!]
\centering
\includegraphics[scale=.5]{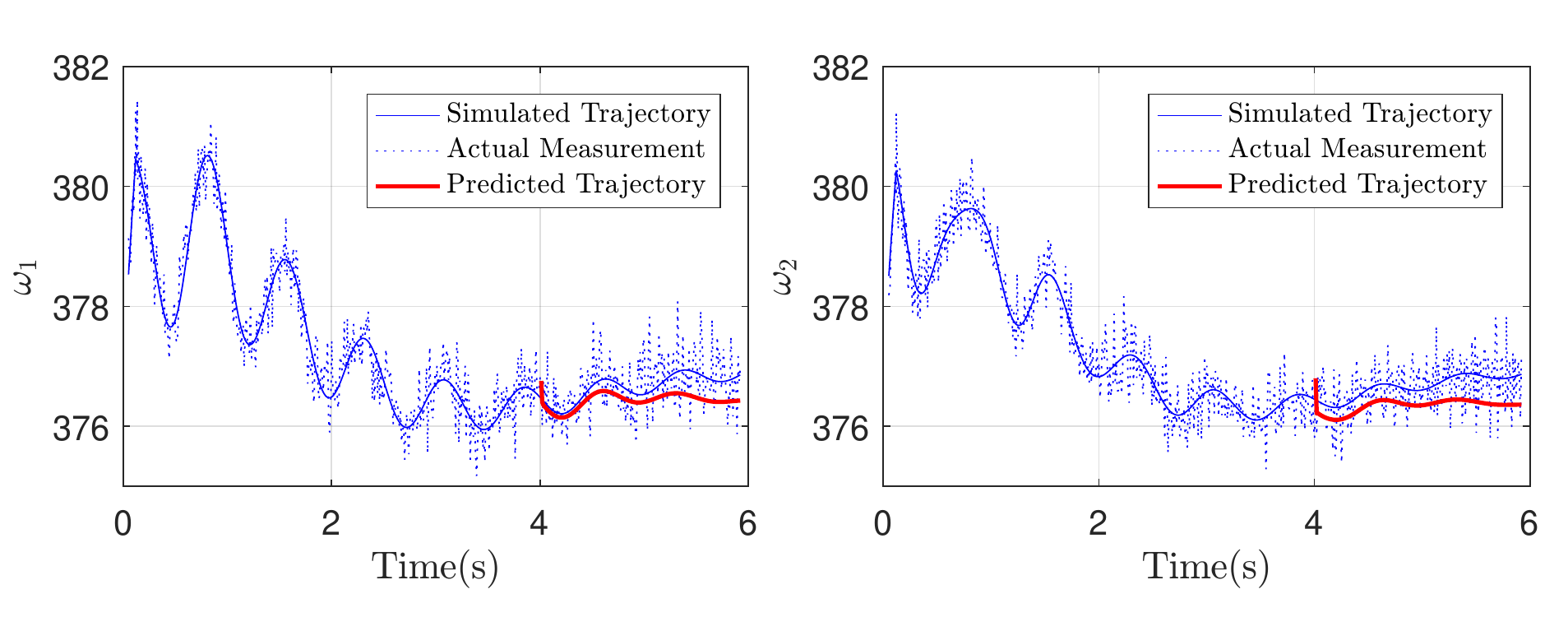}
\caption{Generator Angular Speed prediction for noisy measurement}\label{fig_prediction_with_noise}
\end{figure}
\begin{figure}[htp!]
\centering
\includegraphics[scale=.5]{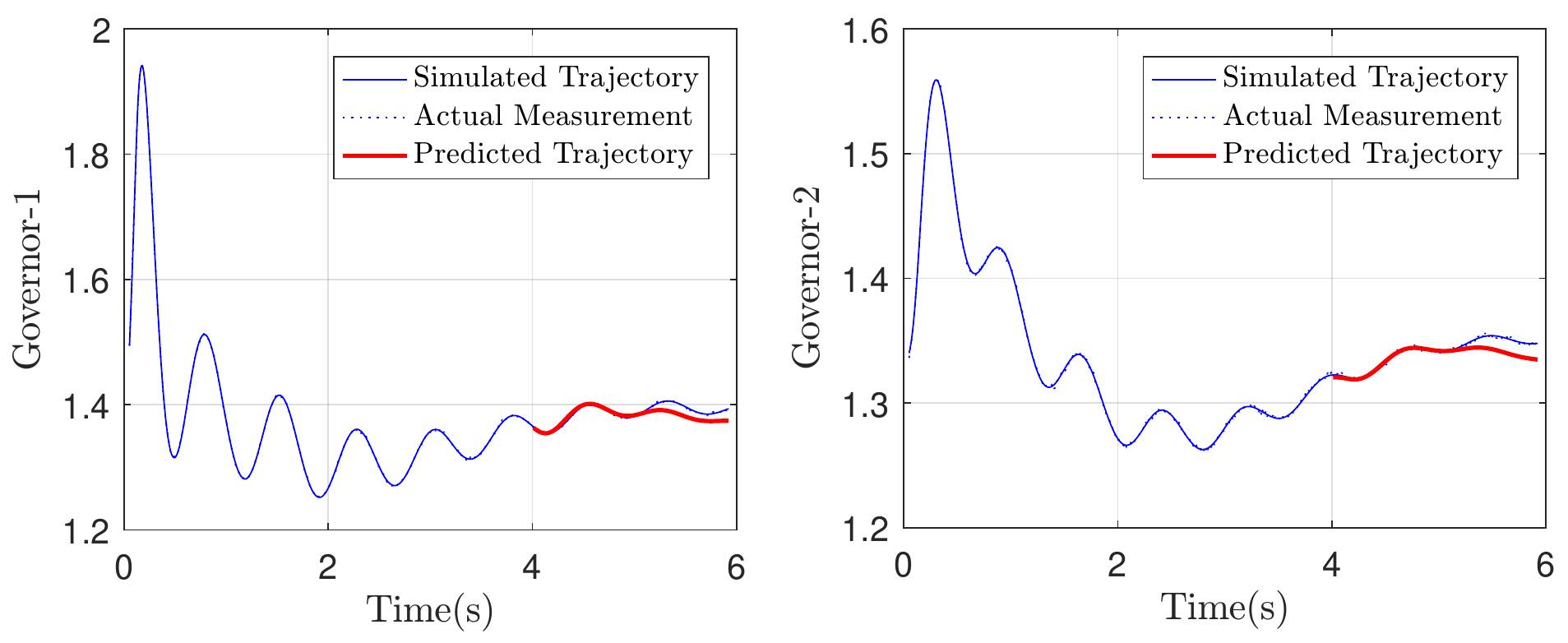}
\caption{Governor Controller state prediction for noisy measurement}\label{fig_prediction_with_noise2}
\end{figure}

As shown in figure \ref{fig_prediction_with_noise}, generator angle are recorded for first 4 seconds and their evolution for next two seconds is predicted. Here the robust predictor is trained over the noisy data (blue dotted line) without any knowledge of system model. The solid blue line in the figure corresponds to simulated data without any noise. We observe that the predicted trajectory (red curve) can estimate the underlying system states from the noisy measurement. Similarly, as shown in figure \ref{fig_prediction_with_noise2}, robust estimator can predict controller and governor states of the system with high accuracy.  
\subsection{Effect of measurement window size}
\begin{figure}[htp!]
\centering
\includegraphics[scale=.6]{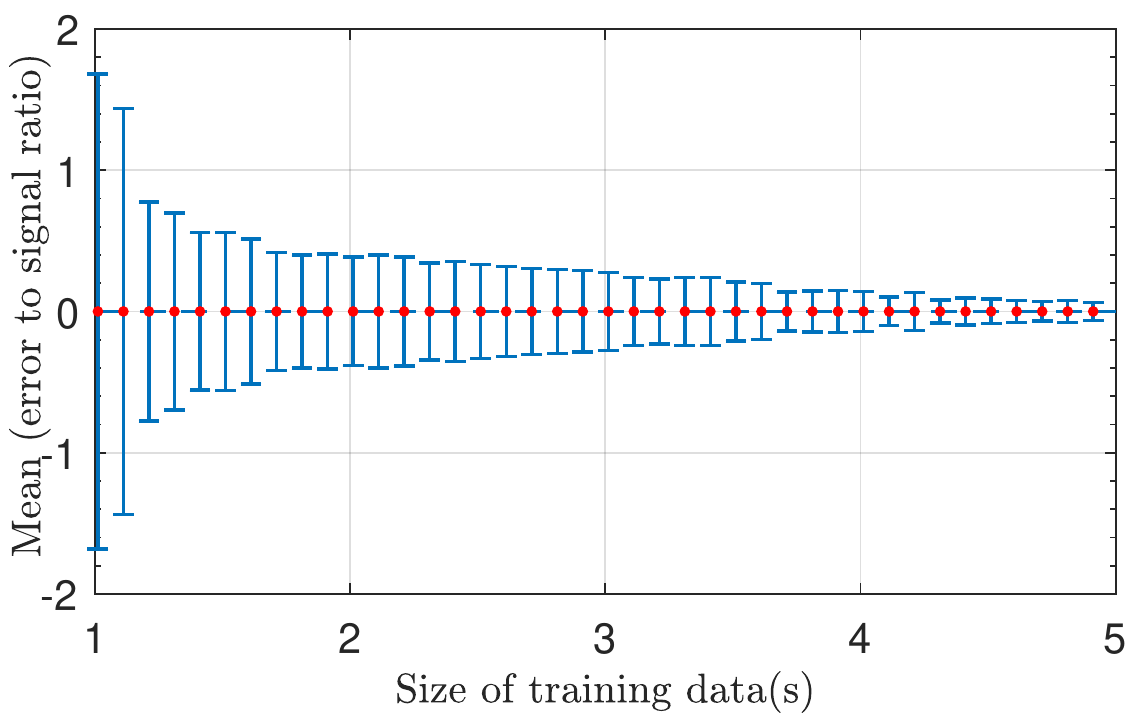}
\caption{Mean trajectory prediction error for all states with respect to length of training data for one second horizon}\label{fig_prediction_error}
\end{figure}

For the given test system, we further predict all state trajectories using robust predictor. It is important to understand the role of length of training data for nonlinear system prediction. For all training samples, next $1 second$ trajectory is predicted for the noisy measurements. Further error is computed with respect to simulated trajectory of deterministic case. As this is the case of post fault oscillations which settles down in seconds, one second prediction is sufficient for dynamic analysis. As shown in figure \ref{fig_prediction_error}, mean error in prediction is computed against the length of training data. It is quiet intuitive that for small data size, prediction will have large error. As visible in figure \ref{fig_prediction_error}, the prediction error is within $5\%$ of mean value for sample size greater than $4$ seconds. Thus for realistic power system scenarios, robust predictor can estimate state trajectories well within 5 seconds of any perturbation. 

\section{Conclusion}\label{section_conclusion}
In this work, we proposed a robust algorithm for approximation of Koopman operator for the identification of nonlinear power system dynamics. This robust approximation approach gives  us the ability to identify system dynamics from a noisy stochastic measurements. Further, theory of robust predictor is also proposed, which enable us to predict true system trajectories in midst of noisy measurements. IEEE 9 bus system is considered to illustrate the application of robust estimation for power system. With realistic noise level of 30-17 dB, robust EDMD was able to identify system dominant mode with great accuracy as oppose to existing methods of linear operator computation. Further, we have shown that robust predictor can estimate nonlinear state trajectories for power system from corrupted measurements. Thus, we can capture the underlying dynamic system for a noisy measurement. We have also tested the accuracy of robust predictor with respect to size of measurement data. Robust predictor is able to estimate system states within a bound of $5\%$ just with a sample size of four seconds. Thus, this robust predictor can be used to identify various dynamic events associated with power system.

\bibliographystyle{IEEEtran}
\bibliography{ref,ref1,reference,ref_new}

\begin{thebibliography}{10}
\providecommand{\url}[1]{#1}
\csname url@samestyle\endcsname
\providecommand{\newblock}{\relax}
\providecommand{\bibinfo}[2]{#2}
\providecommand{\BIBentrySTDinterwordspacing}{\spaceskip=0pt\relax}
\providecommand{\BIBentryALTinterwordstretchfactor}{4}
\providecommand{\BIBentryALTinterwordspacing}{\spaceskip=\fontdimen2\font plus
\BIBentryALTinterwordstretchfactor\fontdimen3\font minus
  \fontdimen4\font\relax}
\providecommand{\BIBforeignlanguage}[2]{{%
\expandafter\ifx\csname l@#1\endcsname\relax
\typeout{** WARNING: IEEEtran.bst: No hyphenation pattern has been}%
\typeout{** loaded for the language `#1'. Using the pattern for}%
\typeout{** the default language instead.}%
\else
\language=\csname l@#1\endcsname
\fi
#2}}
\providecommand{\BIBdecl}{\relax}
\BIBdecl

\bibitem{uhlen2016synchrophasor}
K.~Uhlen, P.~Overholt, and O.~Valentine, ``Synchrophasor applications for wide
  area monitoring and control,'' \emph{ISGAN annex}, vol.~6, 2016.

\bibitem{schweitzer2008advanced}
E.~O. Schweitzer, D.~Whitehead, A.~Guzman, Y.~Gong, and M.~Donolo, ``Advanced
  real-time synchrophasor applications,'' in \emph{proceedings of the 35th
  Annual Western Protective Relay Conference, Spokane, WA}, 2008.

\bibitem{de2010synchronized}
J.~De~La~Ree, V.~Centeno, J.~S. Thorp, and A.~G. Phadke, ``Synchronized phasor
  measurement applications in power systems,'' \emph{IEEE Transactions on smart
  grid}, vol.~1, no.~1, pp. 20--27, 2010.

\bibitem{yin2015data}
S.~Yin, X.~Li, H.~Gao, and O.~Kaynak, ``Data-based techniques focused on modern
  industry: An overview,'' \emph{IEEE Transactions on Industrial Electronics},
  vol.~62, no.~1, pp. 657--667, 2015.

\bibitem{larsson2009monitoring}
M.~Larsson and D.~S. Laila, ``Monitoring of inter-area oscillations under
  ambient conditions using subspace identification,'' in \emph{Power \& Energy
  Society General Meeting, 2009. PES'09. IEEE}.\hskip 1em plus 0.5em minus
  0.4em\relax IEEE, 2009, pp. 1--6.

\bibitem{messina2006interpretation}
A.~Messina, V.~Vittal, D.~Ruiz-Vega, and G.~Enr{\'\i}quez-Harper,
  ``Interpretation and visualization of wide-area pmu measurements using
  hilbert analysis,'' \emph{IEEE Transactions on Power Systems}, vol.~21,
  no.~4, pp. 1763--1771, 2006.

\bibitem{liu2008oscillation}
G.~Liu and V.~Venkatasubramanian, ``Oscillation monitoring from ambient pmu
  measurements by frequency domain decomposition,'' in \emph{Circuits and
  Systems, 2008. ISCAS 2008. IEEE International Symposium on}.\hskip 1em plus
  0.5em minus 0.4em\relax IEEE, 2008, pp. 2821--2824.

\bibitem{susuki2011nonlinear}
Y.~Susuki and I.~Mezic, ``Nonlinear koopman modes and coherency identification
  of coupled swing dynamics,'' \emph{IEEE Transactions on Power Systems},
  vol.~26, no.~4, pp. 1894--1904, 2011.

\bibitem{susuki2014nonlinear}
Y.~Susuki and I.~Mezi{\'c}, ``Nonlinear koopman modes and power system
  stability assessment without models,'' \emph{IEEE Transactions on Power
  Systems}, vol.~29, no.~2, pp. 899--907, 2014.

\bibitem{susuki2016applied}
Y.~Susuki, I.~Mezic, F.~Raak, and T.~Hikihara, ``Applied koopman operator
  theory for power systems technology,'' \emph{Nonlinear Theory and Its
  Applications, IEICE}, vol.~7, no.~4, pp. 430--459, 2016.

\bibitem{budivsic2012applied}
M.~Budi{\v{s}}i{\'c}, R.~Mohr, and I.~Mezi{\'c}, ``Applied koopmanism,''
  \emph{Chaos: An Interdisciplinary Journal of Nonlinear Science}, vol.~22,
  no.~4, p. 047510, 2012.

\bibitem{williams2015data}
M.~O. Williams, I.~G. Kevrekidis, and C.~W. Rowley, ``A data--driven
  approximation of the koopman operator: Extending dynamic mode
  decomposition,'' \emph{Journal of Nonlinear Science}, vol.~25, no.~6, pp.
  1307--1346, 2015.

\bibitem{barocio2015dynamic}
E.~Barocio, B.~C. Pal, N.~F. Thornhill, and A.~R. Messina, ``A dynamic mode
  decomposition framework for global power system oscillation analysis,''
  \emph{IEEE Transactions on Power Systems}, vol.~30, no.~6, pp. 2902--2912,
  2015.

\bibitem{raak2016data}
F.~Raak, Y.~Susuki, and T.~Hikihara, ``Data-driven partitioning of power
  networks via koopman mode analysis,'' \emph{IEEE Transactions on Power
  Systems}, vol.~31, no.~4, pp. 2799--2808, 2016.

\bibitem{susuki2018estimation}
Y.~Susuki, R.~Hamasaki, and A.~Ishigame, ``Estimation of power system inertia
  using nonlinear koopman modes,'' \emph{arXiv preprint arXiv:1805.01967},
  2018.

\bibitem{mauroy2016global}
A.~Mauroy and I.~Mezi{\'c}, ``Global stability analysis using the
  eigenfunctions of the koopman operator,'' \emph{IEEE Transactions on
  Automatic Control}, vol.~61, no.~11, pp. 3356--3369, 2016.

\bibitem{martin2008exploring}
K.~Martin, D.~Hamai, M.~Adamiak, S.~Anderson, M.~Begovic, G.~Benmouyal,
  G.~Brunello, J.~Burger, J.~Cai, B.~Dickerson \emph{et~al.}, ``Exploring the
  ieee standard c37. 118--2005 synchrophasors for power systems,'' \emph{IEEE
  transactions on power delivery}, vol.~23, no.~4, pp. 1805--1811, 2008.

\bibitem{phadke2009synchronized}
A.~G. Phadke and B.~Kasztenny, ``Synchronized phasor and frequency measurement
  under transient conditions,'' \emph{IEEE Transactions on Power Delivery},
  vol.~24, no.~1, pp. 89--95, 2009.

\bibitem{becejac2016analysis}
T.~Becejac, P.~Dehghanian, and M.~Kezunovic, ``Analysis of pmu algorithm errors
  during fault transients and out-of-step disturbances,'' in \emph{Transmission
  \& Distribution Conference and Exposition-Latin America (PES T\&D-LA), 2016
  IEEE PES}.\hskip 1em plus 0.5em minus 0.4em\relax IEEE, 2016, pp. 1--6.

\bibitem{sinha2018robust}
S.~Sinha, B.~Huang, and U.~Vaidya, ``Robust approximation of koopman operator
  and prediction in random dynamical systems,'' in \emph{2018 Annual American
  Control Conference (ACC)}.\hskip 1em plus 0.5em minus 0.4em\relax IEEE, 2018,
  pp. 5491--5496.

\bibitem{singh2014decentralized}
A.~K. Singh, B.~C. Pal \emph{et~al.}, ``Decentralized dynamic state estimation
  in power systems using unscented transformation,'' \emph{IEEE Trans. Power
  Syst}, vol.~29, no.~2, pp. 794--804, 2014.

\bibitem{netto2018robust}
M.~Netto and L.~Mili, ``A robust data-driven koopman kalman filter for power
  systems dynamic state estimation,'' \emph{IEEE Transactions on Power
  Systems}, vol.~33, no.~6, pp. 7228--7237, 2018.

\bibitem{2018arXiv180308562S}
S.~{Sinha}, H.~{Bowen}, and U.~{Vaidya}, ``{On Robust Computation of Koopman
  Operator and Prediction in Random Dynamical Systems},'' \emph{ArXiv
  e-prints}, Mar. 2018.

\bibitem{caramanis201214}
C.~Caramanis, S.~Mannor, and H.~Xu, ``Robust optimization in machine
  learning,'' \emph{Optimization for machine learning}, p. 369, 2012.

\bibitem{Sauer_pai_book}
P.~W. Sauer and M.~Pai, ``Power system dynamics and stability,'' \emph{Urbana},
  vol.~51, p. 61801, 1997.

\bibitem{khaitan2013high}
S.~K. Khaitan and J.~D. McCalley, ``High performance computing for power system
  dynamic simulation,'' in \emph{High Performance Computing in Power and Energy
  Systems}.\hskip 1em plus 0.5em minus 0.4em\relax Springer, 2013, pp. 43--69.

\bibitem{milano2005open}
F.~Milano, ``An open source power system analysis toolbox,'' \emph{IEEE
  Transactions on Power systems}, vol.~20, no.~3, pp. 1199--1206, 2005.

\bibitem{willems2005note}
J.~C. Willems, P.~Rapisarda, I.~Markovsky, and B.~L. De~Moor, ``A note on
  persistency of excitation,'' \emph{Systems \& Control Letters}, vol.~54,
  no.~4, pp. 325--329, 2005.

\end{thebibliography}

\end{document}